\documentclass[12pt]{article}
\usepackage{latexsym,epsfig,graphicx,a4wide,amssymb}
\newcommand{\La}{\mathcal{L}}
\newcommand{\pa}{\partial}

\newcommand{\pkn}{\partial_\nu}
\newcommand{\pkm}{\partial_\mu}
\def\be{\begin{equation}}
 \def\ee{\end{equation}}
 \def\bea{\begin{eqnarray}}
 \def\eea{\end{eqnarray}}
 \def\nn{\nonumber}

\catcode`\@=11

\@addtoreset{equation}{section}

\def\@normalsize{\@setsize\normalsize{15pt}\xiipt\@xiipt
\abovedisplayskip 14pt plus3pt minus3pt%
\belowdisplayskip \abovedisplayskip
\abovedisplayshortskip  \z@ plus3pt%
\belowdisplayshortskip  7pt plus3.5pt minus0pt}
\def\small{\@setsize\small{13.6pt}\xipt\@xipt
\abovedisplayskip 13pt plus3pt minus3pt%
\belowdisplayskip \abovedisplayskip
\abovedisplayshortskip  \z@ plus3pt%
\belowdisplayshortskip  7pt plus3.5pt minus0pt
\def\@listi{\parsep 4.5pt plus 2pt minus 1pt
            \itemsep \parsep
            \topsep 9pt plus 3pt minus 3pt}}

\def\underline#1{\relax\ifmmode\@@underline#1\else
        $\@@underline{\hbox{#1}}$\relax\fi}
\@twosidetrue \relax

\catcode`@=12

\catcode`\@=11

\def\section{\@startsection{section}{1}{\z@}{3.5ex plus 1ex minus
   .2ex}{2.3ex plus .2ex}{\large\bf}}
\def\ps@headings{\def\@oddfoot{}\def\@evenfoot{}
\def\@oddhead{\hbox{}\hfill
        \makebox[.5\textwidth]{\raggedright\ignorespaces --\thepage{}--
        \hfill }}
\def\@evenhead{\@oddhead}
\def\subsectionmark##1{\markboth{##1}{}}
}

\ps@headings

\catcode`\@=12

\begin{document}

\begin{titlepage}

\begin{centering}
\vspace{1cm}
%i
{\Large {\bf Gravitational Particle Production in Gravity Theories
with Non-minimal
Derivative Couplings }}\\

\vspace{1.5cm}

 {\bf  George Koutsoumbas $^{\sharp}$}, {\bf Konstantinos Ntrekis $^\flat$ }\\
 {\bf Eleftherios Papantonopoulos $^{*}$} \\
 \vspace{.2in}
{\bf  Department of Physics, National Technical University of
Athens, \\
Zografou Campus GR 157 73, Athens, Greece}

 \vspace{.2in}

%\vspace{3mm}

\end{centering}
\vspace{1.5cm}

\begin{abstract}

We study the gravitational production of heavy X-particles of mass
of the order of the inflaton mass, produced after the end of
inflation. We find that, in the presence of a derivative coupling of
the inflaton field or of the X-field to the Einstein tensor, the
number of gravitationally produced particles is suppressed as the
strength of the coupling is increased.

\end{abstract}

\vspace{3.5cm}

\begin{flushleft}

$^{\sharp}~~$ e-mail address: kutsubas@central.ntua.gr \\
$ ^{\flat}~~$ e-mail address: drekosk@central.ntua.gr\\
 $^{*} ~~$ e-mail address: lpapa@central.ntua.gr

\end{flushleft}
\end{titlepage}

\vspace{1.5cm}

%\end{titlepage}

\section{Introduction}

Gravitational particle production (for a review see
\cite{Birrell:1982ix,DeWitt1975}) is a generic mechanism for
quantum fields in a curved spacetime background and are analogs of
particle creation in strong electric fields. This mechanism has
been used to explain the presence of dark matter (DM), which is
believed to constitute most of the mass of the Universe, by
generating superheavy long-lived particles after inflation in the
preheating process \cite{Kuzmin:1998uv,Chung:1998zb}.

The inflationary paradigm is well known and extensively studied (for
a review see \cite{inflation}). During inflation, driven by a scalar
field $\phi$, the Universe expands exponentially solving in this way
the horizon and flatness problems of the standard
 cosmology. The large scale structure of the Universe is
 generated through fluctuations during the inflationary period.
 This fixes the mass of the inflaton field to a value of $m_\phi \sim 10^{13}$ GeV.
 During inflation, the inflaton field
slowly rolls down towards the minimum of its potential. Inflation
ends when the potential energy associated with the inflaton field
becomes comparable to the kinetic energy. When this happens all the
energy of the Universe is contained entirely in the form of coherent
oscillations of the inflaton field around the minimum of its
potential. At that time, through a purely classical mechanism of
parametric resonance \cite{Kofman:1994rk}, studied in detail in
\cite{Khlebnikov:1996zt,Kofman:1997yn}, stable very heavy particles
will be produced in excess. These entities are known as X-particles,
with masses $m_X \gtrsim m_\phi$  and will in general overclose the
Universe. Parametric resonance for X-particles is ineffective if X
is either a fermion field or its coupling to inflaton is small
\cite{Khlebnikov:1996zt}.

Another mechanism of generating heavy DM has been proposed in
\cite{Chung:1998zb}. The DM is produced in the transition between an
inflationary and a matter-dominated universe due to the
``nonadiabatic'' expansion of the background spacetime during the
transition acting on the vacuum quantum fluctuations. It was shown
that for a particular range of the masses of the X-particles the DM
needed for the closure the Universe can be produced gravitationally,
independently of the details of the transition between the
inflationary phase and the matter dominated phase.

In an attempt to explain the spectrum of the highest cosmic rays
the authors in \cite{Kuzmin:1998uv} also proposed the generation
of DM through vacuum fluctuations during inflation. They showed
that to have the right number of gravitationally produced
X-particles their masses should be comparable to the inflaton mass
and  their couplings to the inflaton field should be weak.

In this work we study the gravitational particle production in
gravity theories where the inflaton field couples to the Einstein
tensor and the X-field  in addition to its coupling to the inflaton
field it also couples to the Einstein tensor. We show that these
couplings give a generic suppression mechanism for the production of
the heavy X-particles. As the strength of the coupling to the
Einstein tensor of the inflaton field or of  the X-field is
increased, less particles are produced.

The paper is organized as follows. In Section 2 we review the
formalism of gravitational particle production. In Section 3 we
discuss the inflationary phase allowing the inflaton to couple to
the Einstein tensor. In Section 4 we make a systematic study of
gravitational particle production for a wide range of parameters.
Section 5 contains our conclusions.

\section{Gravitational Particle Production}

In this section we will review the basic formalism of
gravitational particle production in a curved spacetime and we
will derive the Bogolyubov coefficients. We will apply this
formalism to a FRW background and we will extend the formalism to
the case of a scalar field coupled to Einstein tensor.

\subsection{Basic Formalism}

We consider a scalar field $\phi$ with the Lagrange density
\footnote{Throughout the paper we use the ``mostly minus''
convention ($+ - - -$).}
 \be
\La = \sqrt{-g}\big\lbrace
\frac{1}{2}\big[g^{\kappa\lambda}\pa_\kappa\phi\pa_\lambda\phi\big]
-V(\phi)\big\rbrace \label{lag1}~, \ee from which we get the
equation of motion for the scalar field
\be \label{KGVath}
 \square \phi + V_\phi = 0 ~, \ee where \be
\qquad \square \phi = (-g)^{-1/2} \pkm\big[ (-g)^{1/2} g^{\mu\nu}
\pkn\phi \big]~. \ee For  the spacelike hypersurfaces defined by a
constant value for $t$ we define the inner product of the solutions
of (\ref{KGVath}) as
 \be \label{es_gin} (\phi_1,\phi_2) \equiv i
\int (-g)^{1/2} g^{0\nu} \big(\phi_1^*(x) \pkn\phi_2(x) -
\phi_2(x) \pkn \phi_1^*(x)\big) d^3x~. \ee We expand the field
 $\phi$ in terms of the complete set of the modes
$\chi_{\vec{k}}$
  \be \hat{ \phi}(x) =
\sum_{\vec{k}}\big(\hat{a}_{\vec{k}} \chi_{\vec{k}}(x) +
\hat{a}^\dag_{\vec{k}} \chi^*_{\vec{k}}(x) \big)~, \label{phimodes}
\ee with $\chi_{\vec{k}}$ satisfying \be (\chi_{\vec{k}},
\chi_{\vec{k}'}) = \delta_{\vec{k}\vec{k}'}~, \qquad
(\chi_{\vec{k}}^*,\chi_{\vec{k}'}^*) = -\delta_{\vec{k}\vec{k}'}~,
\qquad (\chi_{\vec{k}}, \chi_{\vec{k}'}^*) = 0~. \ee The operators
$\hat{a}_{\vec{k}}$ and $\hat{a}^\dag_{\vec{k}}$ satisfy the
commutation relations
 \be [\hat{a}_{\vec{k}} ,
\hat{a}_{\vec{k}'}] = 0~, \qquad [\hat{a}^\dag_{\vec{k}} ,
\hat{a}^\dag_{\vec{k}'}] = 0~, \qquad [\hat{a}_{\vec{k}} ,
\hat{a}^\dag_{\vec{k}'}] = \delta_{\vec{k}{\vec{k}'}}~. \ee The
$n_{\vec{k}}$ excitations are given by \be |n_{\vec{k}}\rangle =
\frac{1}{\sqrt{n_{\vec{k}}!}}(\hat{a}^\dag_{\vec{k}})^{n_{\vec{k}}}|0_\chi\rangle~.
\ee Consider now another complete set of modes $\psi_{\vec{k}'}$.
The transformation connecting the two sets of modes
$\chi_{\vec{k}}$ and $\psi_{\vec{k}'}$ \be \chi_{\vec{k}}(x) =
\sum_{\vec{k}'}\big(\alpha_{\vec{k}{\vec{k}'}} \psi_{\vec{k}'}(x)
+ \beta_{\vec{k}{\vec{k}'}}\psi^*_{\vec{k}'}(x)\big) \ee is the
Bogolyubov transformation and the $\alpha_{\vec{k}{\vec{k}'}} \;,
\; \beta_{\vec{k}{\vec{k}'}}$ are the Bogolyubov coefficients.

If we expand the field $\phi$ also in terms of the modes
$\psi_{\vec{k}'}$ \be \hat{\phi}(x) =
\sum_{\vec{k}'}\big(\hat{b}_{\vec{k}'} \psi_{\vec{k}'}(x) +
\hat{b}^\dag_{\vec{k}'} \psi^*_{\vec{k}'}(x) \big)~, \ee then one
can prove that: \bea \label{btel} \hat{a}_{\vec{k}} &=&
\sum_{\vec{k}'}\big(\alpha^*_{\vec{k}{\vec{k}'}} \hat{b}_{\vec{k}'}
- \beta^*_{\vec{k}{\vec{k}'}} \hat{b}^\dag_{\vec{k}'}\big)~,  \\
\hat{b}_{\vec{k}'} &=& \sum_{\vec{k}}\big(\alpha_{\vec{k}{\vec{k}'}}
\hat{a}_{\vec{k}} + \beta^*_{\vec{k}{\vec{k}'}}
\hat{a}^\dag_{\vec{k}}\big)~. \nn \eea

Imagine that an observer is in the $|0_\psi\rangle$ vacuum where
there are no $\psi$ particles. Using (\ref{btel}) we can calculate
the average number of $\chi$ particles \be \langle 0_\psi |
\hat{n}_{\chi_{\vec{k}}} |0_\psi\rangle = \sum_{\vec{k}'}
|\beta_{\vec{k}{\vec{k}'}}|^2~. \ee Therefore knowledge of the
Bugolyubov coefficients $\beta_{\vec{k}{\vec{k}'}}$ is essential in
order to calculate the $\chi$ particles produced.

\subsection{Particle Production in a FRW Universe}
\label{ssecb}

Consider a 4-dimensional Friedmann-Robertson-Walker Universe with
metric \be ds^2 = dt^2 - a^2(t)d\vec{x}^2~,\label{frwmetric} \ee
where $a(t)$ is the scale factor. Then equation (\ref{KGVath}) for
the scalar field becomes \be \ddot{\phi}(x)  +
3\frac{\dot{a}(t)}{a(t)} \dot{\phi}(x) - a^{-2}(t) \sum^3_{i=1}
\pa_i^2 \phi(x) + V_\phi = 0~.\label{kg1}\ee We consider the
potential function $V(\phi) = \frac{1}{2}\big[ m_\phi^2 \phi^2 +
\zeta R(t) \phi^2 \big],$ where we have included a coupling term of
the scalar field $\phi$ to the curvature $R$. Then (\ref{kg1})
becomes \be \label{meproth} \ddot{\phi}(x)  +
3\frac{\dot{a}(t)}{a(t)} \dot{\phi}(x) - a^{-2}(t) \sum^3_{i=1}
\pa_i^2 \phi(x) + [m_\phi^2 + \zeta R(t)]\phi  = 0~. \ee We expand
$\phi$ in modes as in (\ref{phimodes}) with $\chi_{\vec{k}} \sim
e^{i\vec{k}\vec{x}} \chi_{\vec{k}}(t)$. Then (\ref{meproth}) becomes
\be \ddot{\chi_{\vec{k}}}(t) + 3\frac{\dot{a}(t)}{a(t)}
\dot{\chi_{\vec{k}}}(t)
 +\Big[ \frac{k^2}{a^2(t)}  + m_\phi^2 + \zeta R(t)\Big]  \chi_{\vec{k}}(t)  =
 0~.\label{xmeproth}\ee
With the transformation $\chi_{\vec{k}}(t)=f(t)h_{\vec{k}}(t),$
where $ f(t)=1/a^{2/3}(t),$ we can eliminate the first derivative in
(\ref{xmeproth}) and we obtain \be \ddot{h}_{\vec{k}}(t) + \Big[
m^2_\phi + \frac{k^2}{a^2(t)} + \zeta R(t) -
\frac{3}{2}\frac{\ddot{a}(t)}{a(t)} - \frac{3}{4}
\frac{\dot{a}^2(t)}{a^2(t)} \Big] h_{\vec{k}}(t) =
0\label{xorisproth}~. \ee We assume that asymptotically the
spacetime is Minkowski and define two complete sets of modes for the
scalar field $\phi$, \be \chi^{in}_{\vec{k}}(x) =
\frac{e^{i\vec{k}\vec{x}} h_{\vec{k}}^{in}(t)}{V^{1/2} a(t)^{3/2}}~,
\qquad \chi^{out}_{\vec{k}}(x) = \frac{e^{i\vec{k}\vec{x}}
h_{\vec{k}}^{out}(t)}{V^{1/2}a(t)^{3/2}}~,\label{def1} \ee with
$h_{\vec{k}}^{in}(t)$ and $h_{\vec{k}}^{out}(t)$ solving
(\ref{xorisproth}) with the relevant boundary conditions, and the
volume $V$ appears in (\ref{def1}) to get the correct commutation
relations of the $h_{\vec{k}}$ modes in the case of
$\vec{k}=\vec{k}'$, so that the modes are orthonormal.

Following the formalism of Sec.  2.1  the average number of
$\chi^{in}$ particles can be calculated from \be \label{part1}
\langle 0_{\chi^{out}} | \hat{n}_{\chi^{in}_{\vec{k}}} |
0_{\chi^{out}}\rangle = \sum_{\vec{k}'}
|\beta_{\vec{k}{\vec{k}'}}|^2 = \sum_{\vec{k}'}
|\beta_{\vec{k}}\delta_{-\vec{k}{\vec{k}'}}|^2 =
|\beta_{\vec{k}}|^2~, \ee where $\beta_{\vec{k}}$ are the
Bogolyubov coefficients given by \be  \label{beta} \beta_{\vec{k}}
= -({\chi^{out}_{-\vec{k}}{}}^*,\chi^{in}_{\vec{k}}) =  i\big( h_{\vec{k}}^{in}(t) \dot{h}_{-\vec{k}}^{out}(t) -
h_{-\vec{k}}^{out}(t) \dot{h}_{\vec{k}}^{in}(t) \big)~. \ee

\subsection{Particle Production from a Scalar Field Coupled to Einstein Tensor}
\label{ssecc}

So far we have considered a quantized  scalar fields minimally
coupled to gravity. We will introduce a coupling of a scalar field
to the Einstein tensor and we will study how the formalism of
gravitational particle production is modified in the presence of
this coupling.

The non-minimal couplings between derivatives of a scalar field and
curvature are  types of scalar-tensor theories
\cite{Amendola:1993uh} in which  the field equations  can be reduced
to  second-order differential equations \cite{Sushkov:2009hk}. This
kind of interaction belongs to a wider class of scalar-tensor
theories having galilean symmetry \cite{Nicolis:2008in}. These
properties of the derivative coupling of the scalar field to the
curvature have triggered the interest of the study of the
cosmological implications of this new type of scalar-tensor theory
\cite{Gao:2010vr,Granda:2009fh,Saridakis:2010mf,Germani:2010gm,Tsujikawa:2012mk}.
Also local black hole solutions were discussed in
\cite{Kolyvaris:2011fk} and applications to holographic
superconductivity were presented in
\cite{Alsup:2012kr,Alsup:2013kda}.

To introduce the coupling of the scalar field to the Einstein tensor
the Lagrangian density (\ref{lag1}) is modified to \be \La =
\sqrt{-g}\Big\lbrace\frac{1}{2}\big[g^{\mu\nu} + \lambda
G^{\mu\nu}\big] \pkm \phi \pkn \phi  - V(\phi)\Big\rbrace~,
\label{lag2} \ee where the coupling $\lambda$ has the units
$[\lambda] = M_{pl}^{-2}$. Then, the equation of motion for the
scalar field (\ref{KGVath}) in the cosmological background
(\ref{frwmetric}) becomes \bea \label{Gekskin} && \Big[1 - 3\lambda
\frac{\dot{a}^2(t)}{a^2(t)}\Big]\ddot{\phi}(x) +
3\Big[\frac{\dot{a}(t)}{a(t)} -
\lambda\Big(\frac{\dot{a}^3(t)}{a^3(t)} + \frac{2\dot{a}(t)
\ddot{a}(t)}{a^2(t)}  \Big)  \Big]\dot{\phi}(x)+ \nn \\
&&\Big[- a^{-2}(t) \sum_{i=1}^3 \pa_i^2\phi(x) +
\lambda\frac{2\ddot{a}(t) a(t) + \dot{a}^2(t)}{a^4(t)} \sum_{i =
1}^3 \pa_i^2 \phi(x)  \Big] + V_\phi  = 0~. \eea

The definition of the inner product for scalar fields which are
coupled non-minimally with the Einstein tensor is a tricky matter.
For our case, which involves a diagonal metric and Einstein tensor,
a hint may come from the conserved current resulting from the
application of the Noether theorem. The relevant symmetry is the
phase transformation $\phi(x)\rightarrow e^{i q \theta} \phi(x),\
\phi^\dagger(x)\rightarrow e^{-i q \theta} \phi^\dagger(x).$ The
resulting current is proportional to the expression: $$i \int
(-g)^{1/2} \left[g^{0\nu} + \lambda G^{0\nu} \right]\big(\phi^*(x)
\pkn\phi(x) - \phi(x) \pkn \phi^*(x)\big) d^3x~.$$ We define the
inner product of two different fields (in constant $t$
hypersurfaces, for instance) to read: \be \label{inner}
(\phi_1,\phi_2) \equiv i \int (-g)^{1/2} \left[g^{0\nu} + \lambda
G^{0\nu} \right]\big(\phi_1^*(x) \pkn\phi_2(x) - \phi_2(x) \pkn
\phi_1^*(x)\big) d^3x~. \ee  One may check that in the specific
model we are dealing with, this form of the inner product does not
depend on the hypersurface chosen. 

Using the same potential as before and expanding the scalar field in
modes of $\chi_{\vec{k}}$ we get \bea \label{Gkin} && \Big[1 -
3\lambda \frac{\dot{a}^2(t)}{a^2(t)}\Big]\ddot{\chi}_{\vec{k}}(t) +
3\Big[\frac{\dot{a}(t)}{a(t)} -
\lambda\Big(\frac{\dot{a}^3(t)}{a^3(t)} + \frac{2\dot{a}(t)
\ddot{a}(t)}{a^2(t)}  \Big)  \Big]\dot{\chi}_{\vec{k}}(t)+ \nonumber
\\ && \Big[\frac{k^2}{a^2(t) }  - \lambda k^2
\frac{2\ddot{a}(t) a(t) + \dot{a}^2(t)}{a^4(t)} +m_\phi^2 + \zeta
R(t) \Big]\chi_{\vec{k}}(t)  = 0~. \eea Following the discussion in
Sec. \ref{ssecb},  after eliminating the first derivatives in
(\ref{Gkin}), we get \be \label{diafG} \ddot{h}_{\vec{k}}(t) +
\Omega_k^2(t) h_{\vec{k}}(t) = 0~, \ee where \be
\Omega_k^2(t)=B(t)-\frac{\dot{A(t)}}{2}-\frac{A^2(t)}{4} \ee and
\bea A(t) &=& \frac{3\Big[\frac{\dot{a}(t)}{a(t)} -
\lambda\Big(\frac{\dot{a}^3(t)}{a^3(t)} + \frac{2\dot{a}(t)
\ddot{a}(t)}{a^2(t)}  \Big)  \Big]}{1 - 3\lambda
\frac{\dot{a}^2(t)}{a^2(t)}}~, \nn \\   B(t) &=&
\frac{\frac{k^2}{a^2(t) }  - \lambda k^2  \frac{2\ddot{a}(t) a(t) +
\dot{a}^2(t)}{a^4(t)} +m_\phi^2 + \zeta R(t) }{1 - 3\lambda
\frac{\dot{a}^2(t)}{a^2(t)}}~. \label{ab}\eea The definition of
the inner product (\ref{inner}) adopted above allows us to normalize
the $\chi_{\vec{k}}$ modes, so that they are orthonormal. The
normalized solutions of (\ref{Gkin}) will have the form \be
\chi_{\vec{k}} (x) = \frac{e^{i\vec{k}\vec{x}}
h_{\vec{k}}(t)}{V^{1/2} \sqrt{a(t)} \sqrt{|3\lambda \dot{a}^2(t) -
a^2(t)|}}~. \ee Then, the average number of particle produced in the
presence of the derivative couplings is given by (\ref{part1}) and
(\ref{beta}).

We will calculate the Bogolyubov coefficients (\ref{beta}) solving
numerically the scalar equation (\ref{Gkin}). We rewrite
(\ref{Gkin}) introducing the dimensionless time $\tau = M_{pl} t$
\bea \label{Gkindim} &&\Big[1 - 3\lambda M^2_{pl}
\frac{\dot{a}^2(\tau)}{a^2(\tau)}\Big]\ddot{\chi}_{\vec{k}}(\tau) +
3\Big[\frac{\dot{a}(\tau)}{a(\tau)} - \lambda
M^2_{pl}\Big(\frac{\dot{a}^3(\tau)}{a^3(\tau)} +
\frac{2\dot{a}(\tau) \ddot{a}(\tau)}{a^2(\tau)}  \Big)
\Big]\dot{\chi}_{\vec{k}}(\tau)+\nn \\
&&\Big[\frac{k^2}{M_{pl}^2 a^2(\tau) }  - \lambda k^2
\frac{2\ddot{a}(\tau) a(\tau) + \dot{a}^2(\tau)}{a^4(\tau)}
+\frac{m_\phi^2}{M_{pl}} + \frac{\zeta R(\tau)}{M_{pl}}
\Big]\chi_{\vec{k}}(\tau)  = 0~.\eea Then (\ref{diafG}) becomes \be
\label{diafGt} \ddot{h}_{\vec{k}}(\tau) + \Omega_k^2(\tau)
h_{\vec{k}}(\tau) = 0~, \ee with  \bea A(\tau) &=&
\frac{3\Big[\frac{\dot{a}(\tau)}{a(\tau)} - \lambda
M_{pl}^2\Big(\frac{\dot{a}^3(\tau)}{a^3(\tau)} +
\frac{2\dot{a}(\tau) \ddot{a}(\tau)}{a^2(\tau)}  \Big)  \Big]}{1 -
3\lambda M_{pl}^2 \frac{\dot{a}^2(\tau)}{a^2(\tau)}}~, \nn \\  B(\tau)
&=& \frac{\frac{k^2}{M^2_{pl} a^2(\tau) }  - \lambda k^2
\frac{2\ddot{a}(\tau) a(\tau) + \dot{a}^2(\tau)}{a^4(\tau)} +
\frac{m_\phi^2}{M_{pl}^2} + \frac{\zeta R(\tau)}{M_{pl}^2} }{1 -
3\lambda M_{pl}^2 \frac{\dot{a}^2(\tau)}{a^2(\tau)}}~. \label{abnew}\eea
If we choose a scale factor of the form $a^2(\tau) = C +
D\tanh\left(\frac{\rho \tau}{M_{pl}}\right)$ the spacetime is
asymptotically Minkowski. The first and second derivatives of the
scale factor and the curvature are asymptotically zero. The initial
and final values of $\Omega$ are \bea \omega_{in} &=&
\sqrt{\frac{k^2}{M_{pl}^2(C-D)} + \frac{m_\phi^2}{M_{pl}^2}}~,\nn \\
\omega_{out} &=& \sqrt{\frac{k^2}{M_{pl}^2(C+D)} +
\frac{m_\phi^2}{M_{pl}^2}}~.\eea We solve numerically the
differential equation (\ref{diafGt}) first with the initial
condition $h_{\vec{k}}^{in}(\tau_-) =
\frac{e^{-i\omega_{in}\tau_-}}{\sqrt{2 \omega_{in}}}$ and then with
the initial condition $h_{\vec{k}}^{out}(\tau_+) =
\frac{e^{-i\omega_{out}\tau_+}}{\sqrt{2 \omega_{out}}}$. The
resulting solutions will give the Bogolyubov coefficients from the
relation (\ref{beta}).

\begin{figure}[h!]
\centering \includegraphics[scale=0.92]{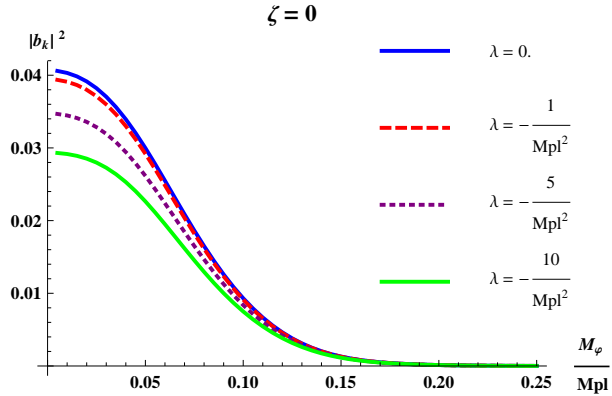} \caption{The
Bogolyubov coefficients  $|\beta_{\vec{k}}|^2$ as a function of
 $m_\phi$ for various values of $\lambda$.} \label{fig:4daz}
\end{figure}

In Fig. \ref{fig:4daz} we show the average number of particles
produced, $\big|\beta_{\vec{k}}|^2,$ in the quantum state with
momentum $\vec{k}.$ In principle the sign of $\lambda$ can be
positive or negative. However, in \cite{Germani:2010gm} stability
arguments were put forward and the avoidance of ghosts restricts
the sign of  $\lambda$ to be negative. Also, an instability was
found in \cite{Kolyvaris:2011fk} looking for local black hole
solutions of a gravity theory in the presence of an
electromagnetic and a scalar field coupled to Einstein tensor with
positive coupling strength $\lambda$. Therefore, for all our
numerics we take $\lambda$ to be negative.

We observe that for a fixed value of the mass of the scalar field,
the number of particles produced  is suppressed as the coupling
$\lambda$ is increased.

In the next sections we will apply this formalism to calculate the
average particle production after the end of inflation in the
preheating epoch. After discussing the inflationary phase of an
expanding Universe with an inflaton field coupled to the Einstein
tensor, in Sec. 4 we will discuss the  particle production of heavy
X-particles.

\section{Inflationary Phase of an Expanding Universe}

In this section after reviewing the basic formalism of the slow roll
inflationary phase we will discuss the inflationary phase in the
presence of the coupling of the inflaton field to the Einstein
tensor.

\subsection{Slow-roll Inflationary Phase}
\label{slrol}

We consider a FRW Universe with a metric (\ref{frwmetric}) and a
scalar field with the action \be S_\phi = \int d^4 x \sqrt{-g} \Big[
\frac{R}{16 \pi G} +\frac{1}{2}g^{\mu\nu}\pkm \phi \pkn \phi -
V(\phi) \Big]~. \label{action2} \ee Varying the above action we get
the Friedmann equation \be \label{Friedmann} H^2(t) =
\frac{8\pi}{3M_{pl}^2}\Big[\frac{\dot{\phi}^2}{2} + V(\phi)\Big]~,
\ee with $H(t) \equiv \frac{\dot{a}(t)}{a(t)}$ and the acceleration
equation \be \label{accel} 2\frac{\ddot{a}(t)}{a(t)} +
\frac{\dot{a}^2(t)}{a^2(t)} =
-\frac{8\pi}{M^2_{pl}}\Big[\frac{\dot{\phi}^2}{2} - V(\phi) \Big]~
\ee while the Klein-Gordon equation for the scalar field $\phi$ is
\be \ddot{\phi}(t)  + 3 H(t) \dot{\phi}(t) + V_\phi =
0~.\label{kleingoerdon} \ee Inflation takes place provided that \be
\left[\ddot{a}(t) > 0\right] \Leftrightarrow
\left[\frac{d}{dt}\Big(\frac{H^{-1}(t)}{a(t)}\Big) < 0\right]  \ee
and using (\ref{accel}) we get the first slow-roll condition
$\frac{\dot{\phi}^2}{2} \ll V(\phi),$ while the second slow-roll
condition is given by $\ddot{\phi} \ll 3H\dot{\phi}$. Using the
slow-roll conditions, equations (\ref{Friedmann}) and
(\ref{kleingoerdon}) become \be H^2 \simeq \frac{8\pi}{3M_{pl}^2}
V(\phi)~, \qquad 3H\dot{\phi} \simeq - V_\phi~, \ee from which we
can find the evolution of the scale factor and the inflaton field
during the inflationary phase \bea a(t) &=& a_0
\exp\Big[-\frac{8\pi}{M_{pl}^2}
\int^{\phi(t)}_{\phi_0} \frac{V}{V_\phi}d\phi\Big]~,\nn \\
\phi(t) &=& \phi_0 - \frac{M_{pl}}{\sqrt{24\pi}} \int^t_{t_0}
\frac{V_\phi}{\sqrt{V}} dt'~, \eea where $a_0=a(t_0) $,
$\phi_0=\phi(t_0)$ and $t_0$ is the  time when the inflation starts.
The inflation ends time $t_f,$ when the kinetic energy becomes
comparable to the potential energy. We may define $t_f$  through the
relation \be \frac{V\big(\phi(t_f)\big)}{V_\phi\big(\phi(t_f) \big)}
= \frac{M_{pl}}{\sqrt{48\pi}}~. \ee

To study the inflationary phase numerically, we fix the potential
to $V(\phi)=\frac{1}{2}M_{\phi}^2 \phi^2$. Then inflation
ends at \be t_{f} = \frac{-M_{pl} + 2 \phi_0 \sqrt{3\pi}}{M_\phi
M_{pl}} \ee
We define a
dimensionless time $\tau \equiv M_{\phi} t$ and a dimensionless
field $\psi(\tau) \equiv \frac{\phi(t)}{M_{pl}}$ so previous relations can be rewritten as \be a(\tau)
= a_0 \exp\Big[\sqrt{\frac{4\pi}{3}} \Big( \psi_0 \tau -
\frac{\tau^2}{\sqrt{48\pi}}\Big) \Big]~, \qquad  \qquad \psi(\tau)
= \psi_0 - \frac{\tau}{\sqrt{12\pi}}~, \ee \be \tau_f =
2\sqrt{3\pi} \psi_0 - 1~, \qquad  \qquad \psi(\tau_f) =
\frac{1}{\sqrt{12\pi}}~, \qquad  \qquad \dot{\psi}(\tau_0) =
-\frac{1}{\sqrt{12\pi}}~. \ee We use the following values for the
numerics \be M_{pl} = 10^{19} GeV \; ,\; \psi_0 = 3.5 \;,\; t_0 =0
\; , \; a_0 = 1 \;  ,\; Mœ_\phi = 10^{-6} M_{pl}~. \label{parm}
\ee The value of $\psi_0 = 3.5$ is fixed from the requirement to
have the right number of e-folds given by the relation \be N
\equiv \ln \frac{a(\tau_f)}{a(\tau_0)}~. \ee

\begin{figure}[h!]
\centering \includegraphics[scale=0.85]{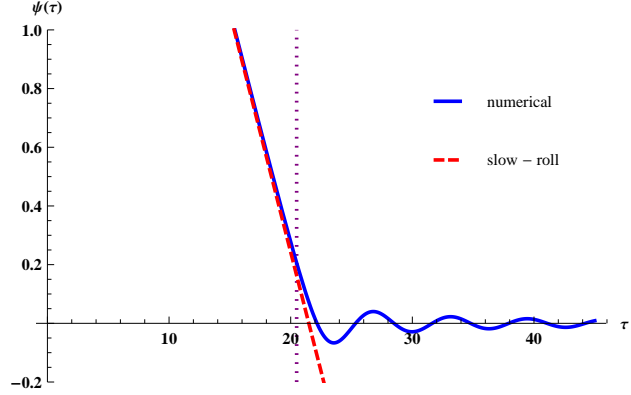} \caption{The
evolution of $\psi$ as function of $\tau$. The continuous line shows
the numerical solution of $\psi$, the dashed line the solution from
the slow-roll approximation, while the vertical line indicates the
end of slow-roll phase. } \label{fig:psin}
\end{figure}

In Fig. \ref{fig:psin} we show the evolution of  $\psi(\tau)$ as it
results from the numerical solution of the initial equations
(\ref{Friedmann}), (\ref{kleingoerdon}) and the slow-roll
approximation. Until the end of inflation the numerical and the
approximate solution coincide, while at the end of the inflation the
$\psi$ field  oscillates at the bottom of the potential producing
particles until it discharges.

\subsection{Inflationary Phase with the Inflaton Field Coupled to
the Einstein Tensor}

In the presence of the derivative coupling of the inflaton field to
the Einstein tensor the action is modified to \be S_\phi = \int d^4
x \sqrt{-g} \Big\lbrace  \frac{ R}{16 \pi G} +
\frac{1}{2}g^{\mu\nu}\pkm \phi \pkn \phi + \frac{1}{2} \lambda_1
G^{\mu\nu}\pkm\phi\pkn\phi - V(\phi)  \Big\rbrace~. \ee We have
denoted by $\lambda_1$ the coupling of the inflaton field to the
Einstein tensor; later on a similar coupling $\lambda_2$ will appear
in connection with the X-field. Varying the above action we get the
Einstein equations \cite{Sushkov:2009hk,Sushkov:2012za}
 \be G_{\mu\nu} = - 8 \pi G
\big[ T_{\mu\nu} + \lambda_1 \Theta_{\mu\nu}\big]~, \ee
 with
  \be
T_{\mu\nu} = \pkm\phi \pkn \phi - \frac{1}{2} g_{\mu\nu}g^{ab}
\pa_a \phi \pa_b \phi + g_{\mu\nu} V(\phi)~, \ee and
 \bea
\Theta_{\mu\nu} = &-&\frac{1}{2} \nabla_\mu  \phi \nabla_\nu \phi R + 2 \nabla_a
\phi \nabla_{(\mu}\phi {R^a_\nu}_)  - \frac{1}{2} G_{\mu\nu}
(\nabla\phi)^2 + \nabla^a \phi \nabla^b\phi R_{\mu a \nu b} +
\nabla_\mu \nabla^a \phi \nabla_\nu \nabla_a \phi \nonumber \\
&-&\nabla_\mu \nabla_\nu \phi \square \phi + g_{\mu\nu} \big[ -
\frac{1}{2}\nabla^a \nabla^b \phi \nabla_a \nabla_b \phi +
\frac{1}{2}(\square \phi)^2 - \nabla_a \phi \nabla_b \phi R^{ab}
\big]~. \eea

In the cosmological background (\ref{frwmetric}) and with a
quadratic potential $V(\phi)=\frac{1}{2}M_{\phi}^2 \phi^2 $ we get
the Friedmann equation \be \frac{\dot{a}^2(t)}{a^2(t)} = \frac{4\pi
}{3M_{pl}^2 }\Big[ \dot{\phi}^2(t)\Big(1 -
9\lambda_1\frac{\dot{a}^2(t)}{a^2(t)}\Big) + M_\phi^2 \phi^2(t)
\Big]~, \ee while the field equation for the scalar field
(\ref{Gekskin}) becomes \be \Big(1-
3\lambda_1\frac{\dot{a}^2(t)}{a^2(t)}\Big) \ddot{\phi}(t) + \Big(3
\frac{\dot{a}(t)}{a(t)} - 3\lambda_1
\Big(\frac{\dot{a}^3(t)}{a^3(t)} +
\frac{2\dot{a}(t)\ddot{a}(t)}{a^2(t)} \Big) \Big) \dot{\phi}(t) +
M_\phi^2 \phi(t)= 0~. \ee Defining as before the dimensionless time
$\tau \equiv M_{\phi} t$, we get \be \label{fried1}
\frac{\dot{a}^2(\tau)}{a^2(\tau)} = \frac{4\pi }{3}\Big[
\dot{\psi}^2(\tau)\Big(1 - 9\lambda_1 M_\phi^2
\frac{\dot{a}^2(\tau)}{a^2(\tau)}\Big) +  \psi^2(\tau)  \Big]~, \ee
\be \label{kg2} \Big(1- 3\lambda_1 M_\phi^2
\frac{\dot{a}^2(\tau)}{a^2(\tau)}\Big) \ddot{\psi}(\tau) + \Big(3
\frac{\dot{a}(\tau)}{a(\tau)} - 3\lambda_1 M_\phi^2
\Big(\frac{\dot{a}^3(\tau)}{a^3(\tau)} +
\frac{2\dot{a}(\tau)\ddot{a}(\tau)}{a^2(\tau)} \Big) \Big)
\dot{\psi}(\tau) + \psi(\tau)= 0~, \ee with $\psi(\tau) \equiv
\phi(\tau)/M_\phi. $

\begin{figure}[h!]
\centering \includegraphics[scale=0.9]{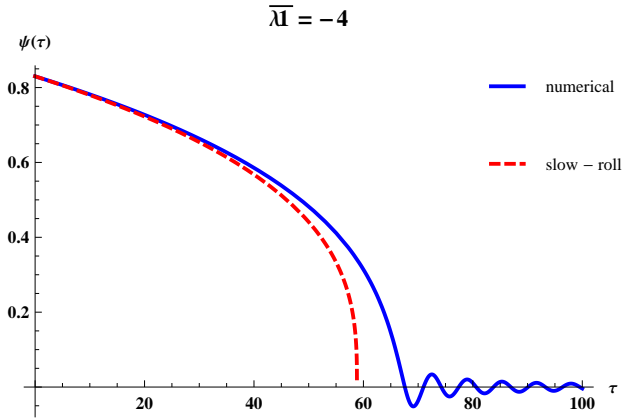}
\caption{Numerical solution of Friedmann and scalar equations
(continuous lines), and the solution of the equations in the slow
roll approximation (dashed lines).} \label{fig:ksi_-4_SR_I}
\end{figure}

We will solve (\ref{fried1}) and (\ref{kg2}) both numerically and in
the slow-roll approximation. For the slow-roll approximation in
addition to the conditions $\frac{\dot{\phi}^2}{2} \ll V(\phi)$ and
 $\ddot{\phi} \ll 3H\dot{\phi}$ we also consider the conditions $3\lambda_1 H^2 \gg 1$
and $\frac{\dot{H}}{H} \ll1$. We define the dimensionless parameter
$\overline{\lambda}_1 \equiv \lambda_1 M_{\phi}^2$. To have the
right number of e-folds ($N\sim 70$), we choose the initial
conditions $\psi(\tau_0)=0.83$, $a(\tau_0)=1$ and
$\overline{\lambda}_1=-4$. Then the time evolution of $\psi(\tau)$
is shown in Fig. \ref{fig:ksi_-4_SR_I}.

\begin{figure}[h!]
\centering \includegraphics[scale=0.9]{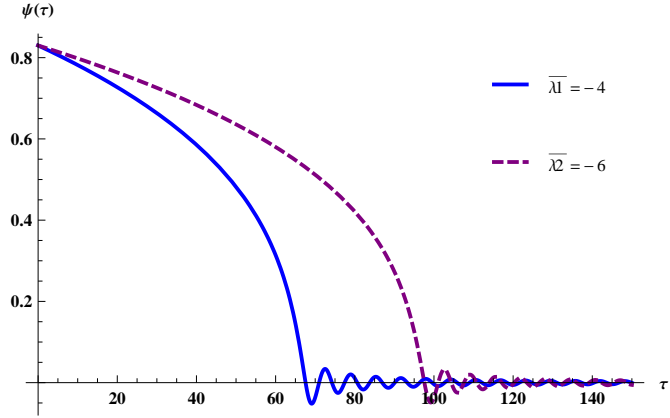}
\caption{Numerical solution of Friedmann and scalar equations for
$\psi(\tau)$ with $\overline{\lambda}_1 = -4$ (continuous lines) and
$\overline{\lambda}_1 = -6$ (dashed lines).} \label{fig:ksi_-4_-6}
\end{figure}

An important observation is that the introduction of the coupling of
the inflaton field to Einstein tensor allows us to go beyond the
Chaotic Inflation Scenario \cite{Linde} in which the inflaton mass
is comparable to the Planck scale. By choosing appropriate values of
the coupling $\lambda_1$ we can have the right number of e-folds for
much lower values of the  $\psi$ field mass. Another observation is
that the coupling to Einstein tensor acts as a friction term,
absorbing energy from the kinetic energy of the inflaton field as it
rolls down the potential, as first noticed in \cite{Germani:2010gm}.
This results in prolonging the duration of the inflation. This
effect is generic and does not depend on the slow roll
approximation, as it can be seen in Fig. \ref{fig:ksi_-4_-6}.

The inflation ends when the kinetic energy becomes comparable to
the potential energy \be\dot{\phi}^2(t) \Big( 1 - 9\lambda_1
\frac{\dot{a}^2(t)}{a^2(t)}\Big) \simeq 2V(\phi)~. \ee This gives
\be \dot{\psi}^2(\tau) =
\frac{\psi^2(\tau)}{1-9\overline{\lambda}_1 H^2(\tau)}~. \label
{rel12}\ee The above relation allows  a very small window of
positive $\overline{\lambda}_1$ \be 0<
\overline{\lambda}_1<\frac{1}{9 H^2(\tau)}~. \label{rel13} \ee We
will discuss its significance in Section 4 where we will present
the gravitational particle production after inflation in the
presence of the derivative coupling.

\section{Particle Production after the End of Inflation}

We will study the gravitational particle production after the end
of inflation using two coupled scalar fields in an expanding
Universe. The scalar fields are the inflaton field $\phi(t)$ which
drives the inflation and a quantum field $X(x)$ which produces the
X-particles. The action of the theory is \bea
S &=&\int d^4 x \sqrt{-g} \frac{R}{16 \pi G} \nn \\
 &+& \int d^4 x \sqrt{-g} \Big\lbrace \frac{1}{2}\Big[\big(g^{\mu\nu} + \lambda_1 G^{\mu\nu}\big)
 \pkm \phi \pkn \phi - M_\phi^2 \phi^2 \Big] \nn \\
 &+& \frac{1}{2}\Big[ \big(g^{\mu\nu} +
 \lambda_2 G^{\mu\nu}\big) \pkm X \pkn X - ( M_X^2 + \zeta R + g^2 \phi^2 )
 X^2\Big]\Big\rbrace~.
\eea The inflaton field couples to  Einstein tensor with field
strength $\lambda_1$, the X-field couples to the Einstein tensor
with field strength $\lambda_2$ and we have also included a direct
coupling of the inflaton to X-field with coupling $g$.

The strategy which we will follow is to find the value of the
inflaton field at the end of inflation with or without its coupling
to Einstein tensor and then through its coupling to the quantum
X-field to calculate the number of X-particles produced due to the
expansion of the Universe \footnote{In \cite{Sadjadi:2012zp}  a
model with derivative couplings was discussed in which the inflaton
field decayed to particles via a phenomenological field.}.

Following the analysis of Sec. \ref{ssecc} we end up to the
differential equation \be \label{Rescdiaf} \ddot{h}_{\vec{k}}(\tau)
+ \Omega_k^2(\tau)  h_{\vec{k}}(\tau) =0~, \ee with \be
\Omega_k^2(\tau) = B(\tau) - \frac{\dot{A}(\tau)}{2} -
\frac{A^2(\tau)}{2} \ee and \bea A(\tau) &=&
3\frac{\frac{\dot{a}(\tau)}{a(\tau)} - \overline{\lambda}_2
\Big(\frac{\dot{a}^3(\tau)}{a^3(\tau)}
 + 2\frac{\dot{a}(\tau)\ddot{a}(\tau)}{a^2(\tau)}\Big)}{1 - 3\overline{\lambda}_2 \frac{\dot{a}^2(\tau)}{a^2(\tau)} }~, \nn \\
 [0.15cm] B(\tau) &=& \frac{\frac{k^2}{M_\phi^2 a^2(\tau)} - \overline{\lambda}_2 k^2 \frac{2a(\tau) \ddot{a}(\tau)
 + \dot{a}^2(\tau)}{a^4(\tau)} + \frac{M_x^2}{M_\phi^2} + \frac{\zeta R(\tau)}{M_\phi^2} +
  \frac{g^2 \psi(\tau) M^2_{pl}}{M_\phi^2 }}{1 - 3\overline{\lambda}_2 \frac{\dot{a}^2(\tau)}{a^2(\tau)}
  }~.
\eea We have used the notations $\tau \equiv M_\phi t,\
\overline{\lambda}_2 \equiv \lambda_2 M_\phi^2.$ After the end of
inflation we assume that the Universe enters a matter domination
epoch with the scale factor $a(\tau)= \tau^{2/3}$. It is more
convenient to calculate the Bogolyubov coefficients from the
differential equations \cite{Chung:1998bt} \bea
\dot{\alpha}_k(\tau) &=& \frac{\dot{\Omega}_k(\tau)}{2\Omega_k(\tau)} \exp\Big[ 2i \int \Omega_k(\tau') d\tau'\Big] \beta_k(\tau)~, \nn \\
\dot{\beta}_k(\tau) &=& \frac{\dot{\Omega}_k(\tau)}{2\Omega_k(\tau)}
\exp\Big[ -2i \int \Omega_k(\tau') d\tau'\Big] \alpha_k(\tau)~. \eea
Then a function of the form \be h_{\vec{k}}(\tau) =
\frac{\alpha(\tau)}{\sqrt{2\Omega_k(\tau)}} e^{-i\int
\Omega_k(\tau') d\tau} + \frac{\beta(\tau)}{\sqrt{2\Omega_k(\tau)}}
e^{i\int \Omega_k(\tau') d\tau} \ee solves (\ref{Rescdiaf}). We will
first calculate the number of particles produced without the
couplings to the Einstein tensor and then we will switch on
$\lambda_1$ and $\lambda_2$ and compare the results.

\subsection{Particle Production without Derivative Couplings}

We consider the usual slow-roll inflationary phase as it was
discussed in Sec. \ref{slrol} without the coupling of the inflaton
field to the Einstein tensor ($\lambda_1=0$). Also the quantum field
X does not couple to Einstein tensor ($\lambda_2=0$). We only allow
the coupling of the X-field to the inflaton field. The inflation
starts at $\tau_0=0$ and with the values of the parameters
(\ref{parm}) ends at $\tau_f=20$. Then the inflaton oscillates
around the minimum of the potential until it discharges. The
Universe enters a matter-dominated epoch and, according to our
discussion in Sec. \ref{ssecb},  X-particles will be produced
\footnote{Particles will also be produced through the resonance
effect. However, we will not discuss this process here.  We also
assumed that there is no particle production during inflation.}.

\begin{figure}[H]
\centering \includegraphics[scale=0.8]{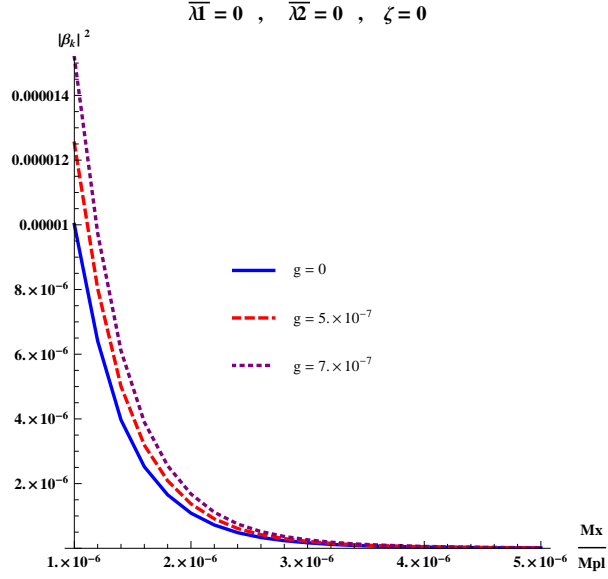} \caption{Bogolyubov
coefficients of the X-field for various values of its coupling to
$\psi$-field.} \label{fig:nogg}
\end{figure}
\begin{figure}[H]
\centering \includegraphics[scale=0.8]{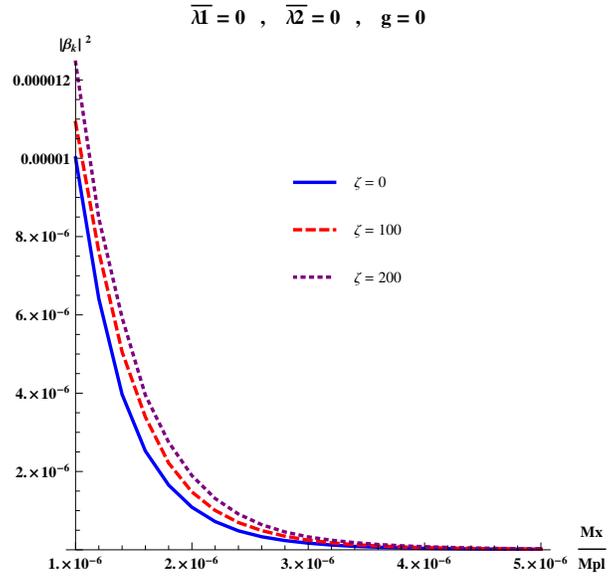}
\caption{Bogolyubov coefficients of the X-field for various values
of its coupling to the curvature. } \label{fig:nogtz}
\end{figure}

We consider the following initial conditions and  values of the
parameters \be \alpha(\tau_f)=1,\,\,\,\, \beta(\tau_f)=0,\,\,\,
M_{\phi}=10^{-6}M_{pl},\,\,\,\,\ k=10^{-5}M_{pl}~. \ee Then we have
$\Omega_k^2(\tau)> 0$ and we know that the adiabatic approximation
$\left(\frac{\dot{\Omega}_k(\tau)}{\Omega_k(\tau)}<<1\right)$ holds
\cite{Parker1969,Parker:2012at}.

In Fig. \ref{fig:nogg}    we show the average number of particles
 $|\beta_{\vec{k}}|^2$ as a function of the mass of the
X-particle $M_X$ for $M_X> 10^{-6}M_{pl}$ for various values of
the coupling $g$ between the inflaton and the X-field and in Fig.
\ref{fig:nogtz} we show the average number of particles
 for various values of the coupling of the X-field to
curvature.

In both cases we observe that for a fixed mass of the X-particle
we have an enhancement of the particle production as the values of
the couplings are increased. Note that in order to have a sizeable
effect in the case that the X-field couples to the curvature, the
coupling $\zeta$ should be large.

\subsection{Particle Production with Derivative Couplings}

We now turn on the derivative couplings. We start with the inflaton
field coupled to Einstein tensor with $\overline{\lambda}_1=-4$ and
the X-field coupled to Einstein tensor. We consider the case in
which there is no coupling between the X-field and the inflaton
field and also there is no coupling to curvature.
\begin{figure}[H]
\centering \includegraphics[scale=0.8]{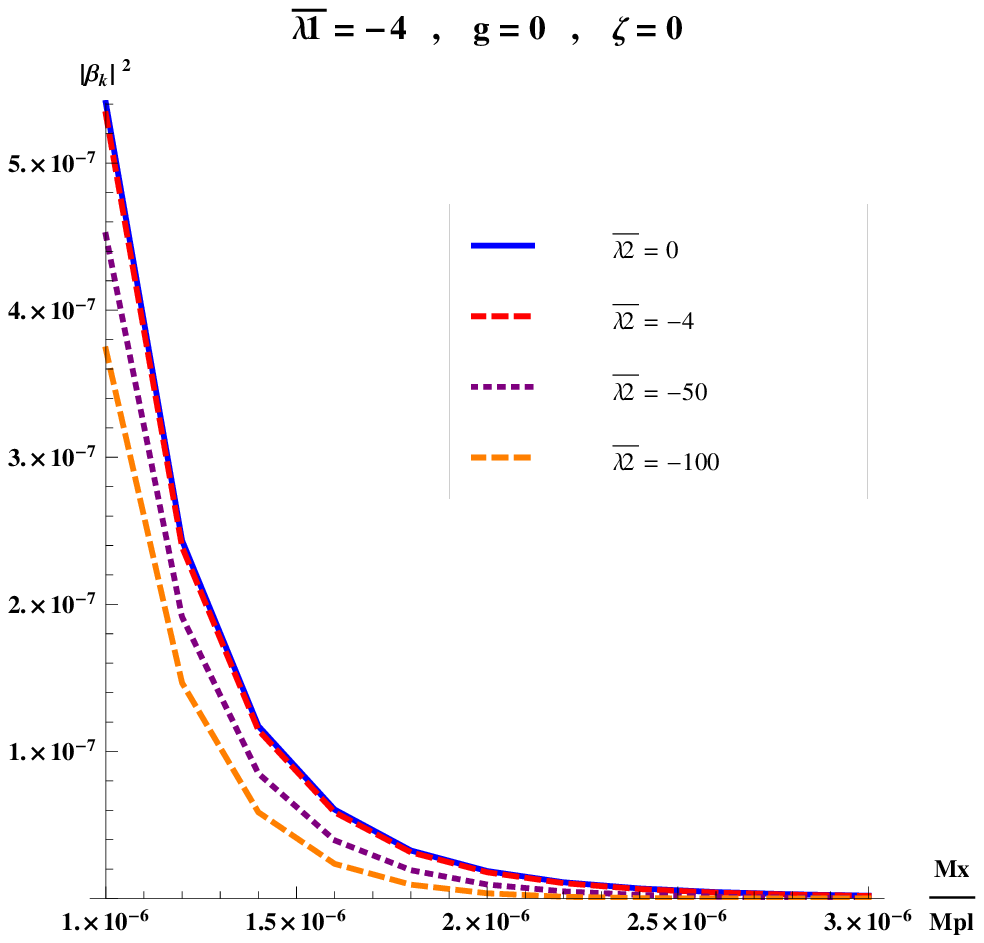}
\caption{Bogolyubov coefficients of the X-field for various values
of its coupling to the Einstein tensor. } \label{fig:G-4ksi}
\end{figure}

In Fig. \ref{fig:G-4ksi} we have the first indication that as the
coupling $\lambda_2$ is increased (in absolute values) less
particles are produced. However, to have a sizable effect the
coupling strength of $\lambda_2$ should be very large, comparable
to $\lambda_1$, in order to compensate the attenuation of the
Einstein tensor due to the expansion of the Universe.

\begin{figure}[H]
\centering \includegraphics[scale=0.8]{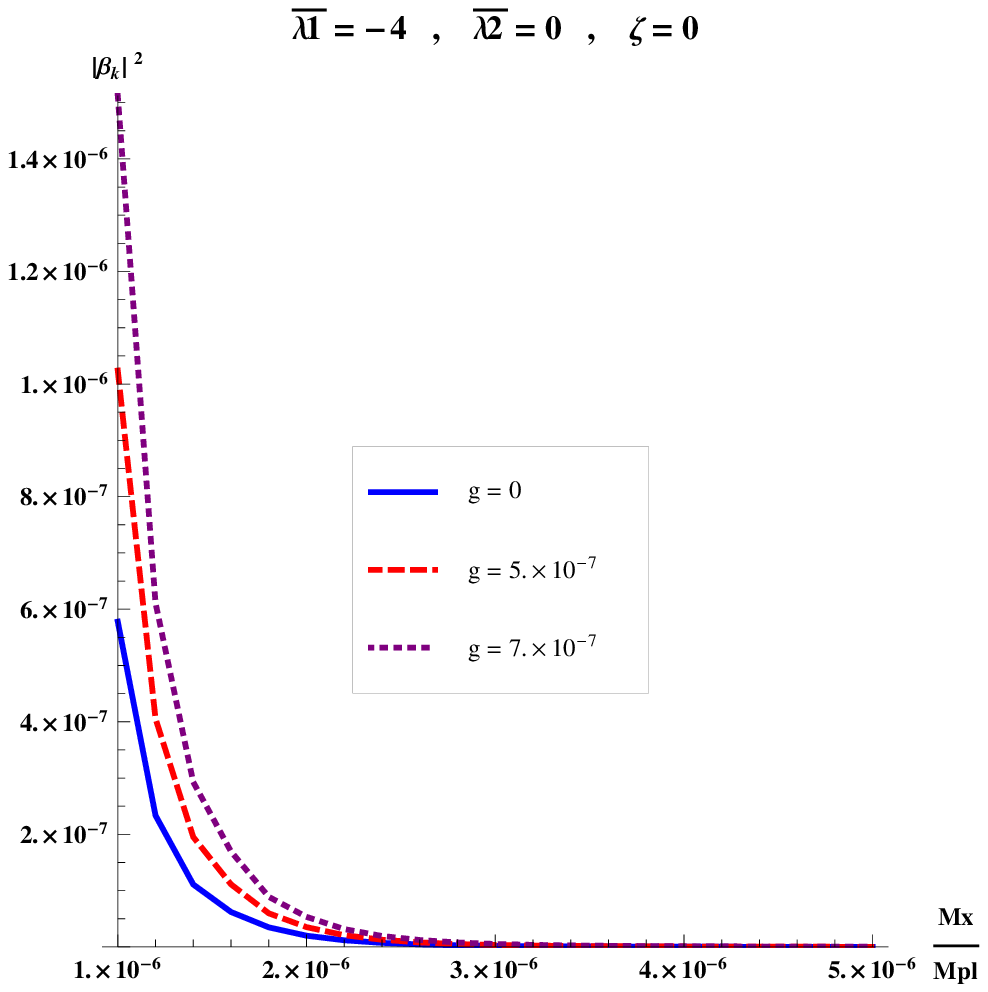}
\caption{Bogolyubov coefficients of the X-field with no couplings
to Einstein tensor, for various values of its coupling to the
inflaton. } \label{fig:ksi-40g}
\end{figure}

\begin{figure}[H]
\centering \includegraphics[scale=0.8]{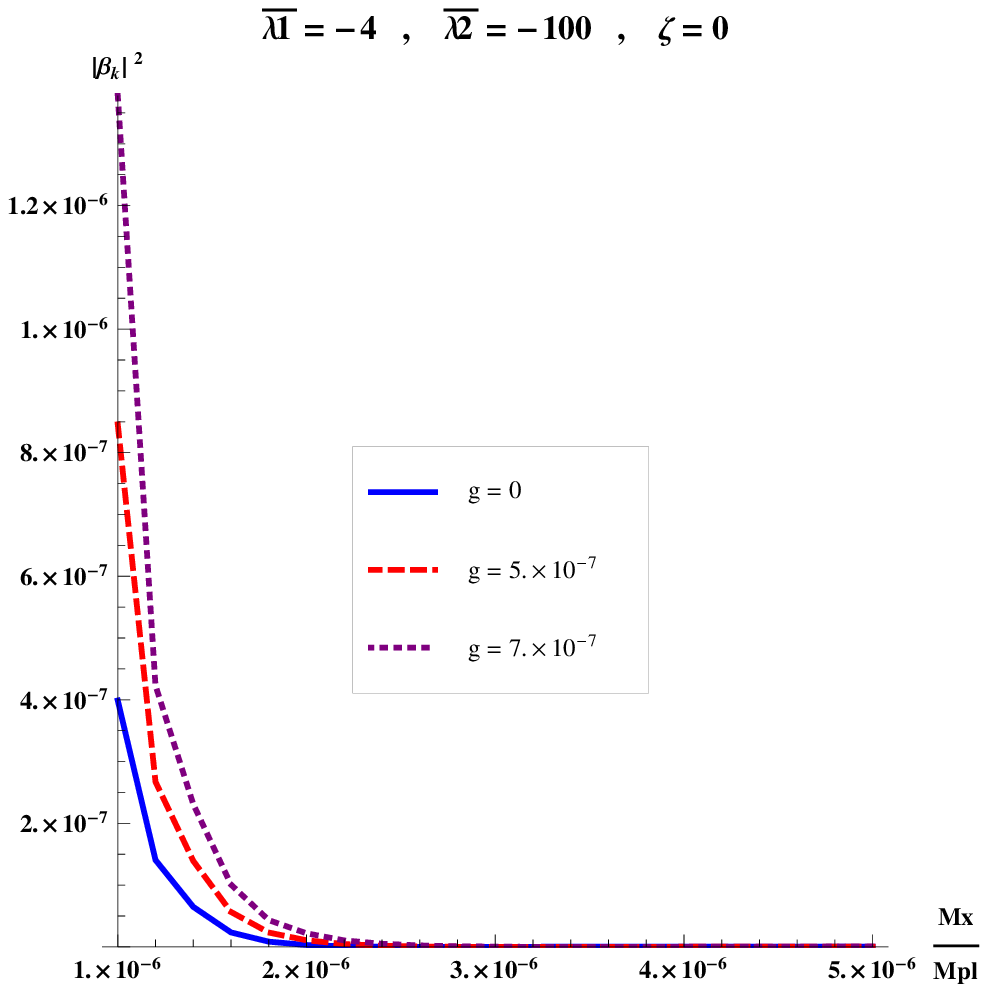}
\caption{Bogolyubov coefficients of the X-field coupled to
Einstein tensor, for various values of its coupling to the
inflaton.  } \label{fig:ksi-4-100g}
\end{figure}

In Fig. \ref{fig:ksi-40g}    and Fig. \ref{fig:ksi-4-100g}  we
show the particles produced if we turn on the coupling of the
inflaton field to X-field. We see that with the increase of the
coupling $g$ we have an increase of the number of particles
produced. However the X-particles produced are less than the
X-particles produced with $\lambda_1=\lambda_2=0$ as can be seen
in Fig. \ref{fig:nogg}.


\begin{figure}[H]
\centering \includegraphics[scale=0.8]{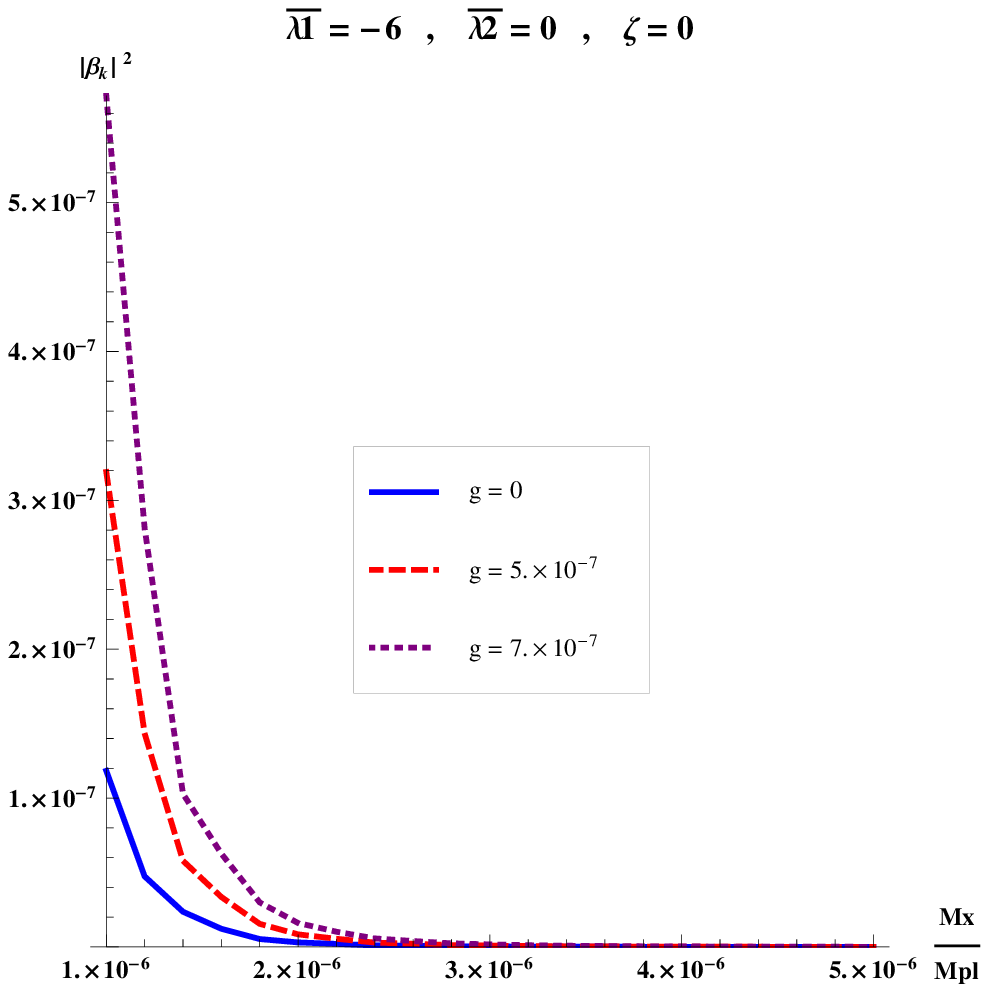}
\caption{Bogolyubov coefficients of the X-field with no couplings
to Einstein tensor, for various values of its coupling to the
inflaton. } \label{fig:ksi-60g}
\end{figure}

\begin{figure}[H]
\centering \includegraphics[scale=0.8]{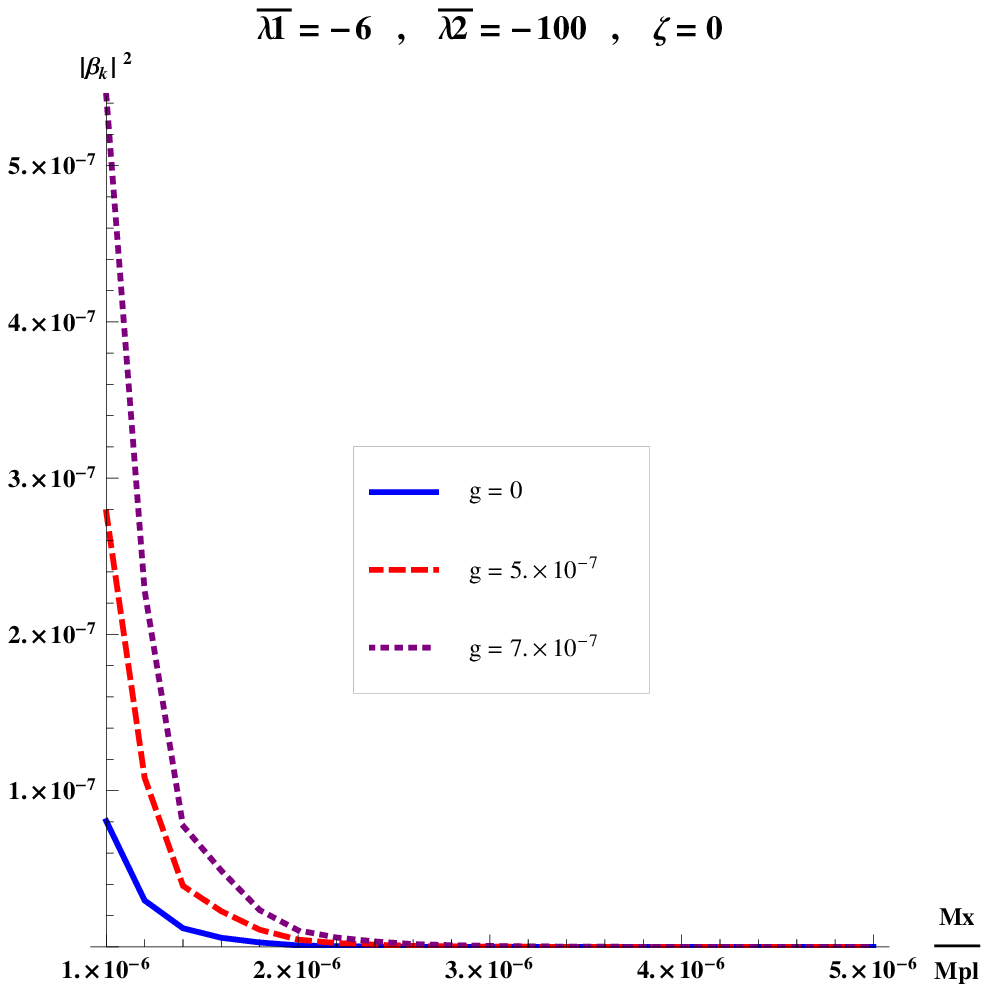}
\caption{Bogolyubov coefficients of the X-field coupled to
Einstein tensor, for various values of its coupling to the
inflaton. } \label{fig:ksi-6-100g}
\end{figure}

\begin{figure}[H]
\centering \includegraphics[scale=0.7]{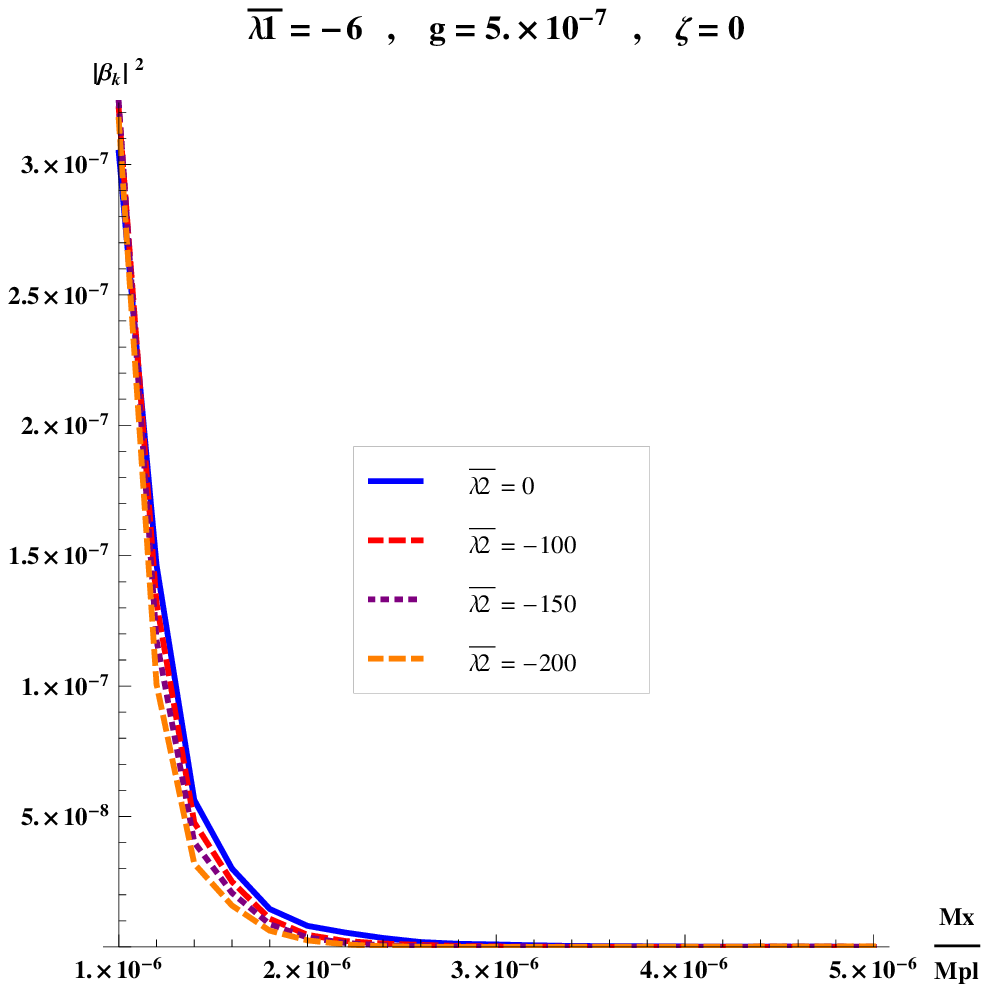}
\caption{Bogolyubov coefficients of the X-field for various values
of its coupling to the Einstein tensor. } \label{fig:ksi-6ksi}
\end{figure}

This effect can be seen more clearly if we increase the coupling of
the inflaton field to Einstein tensor as can be seen in Fig.
\ref{fig:ksi-60g} and  Fig. \ref{fig:ksi-6-100g}. In Fig.
\ref{fig:ksi-6ksi} we fix the value of $g=5.10^{-7}$ and the value
of $\overline{\lambda}_1=-6$ and we vary the value of
$\overline{\lambda}_2$. We see that less particles are produced as
the absolute value of the coupling $\overline{\lambda}_2$ is
increased.


%\begin{figure}[H]
%\centering \includegraphics[scale=0.8]{ksi-6-100g.pdf}
%\caption{Bogolyubov coefficients of the X field for various values
%of its coupling to the Einstein tensor.} \label{fig:ksi-6-100g}
%\end{figure}

\begin{figure}[H]
\centering \includegraphics[scale=0.7]{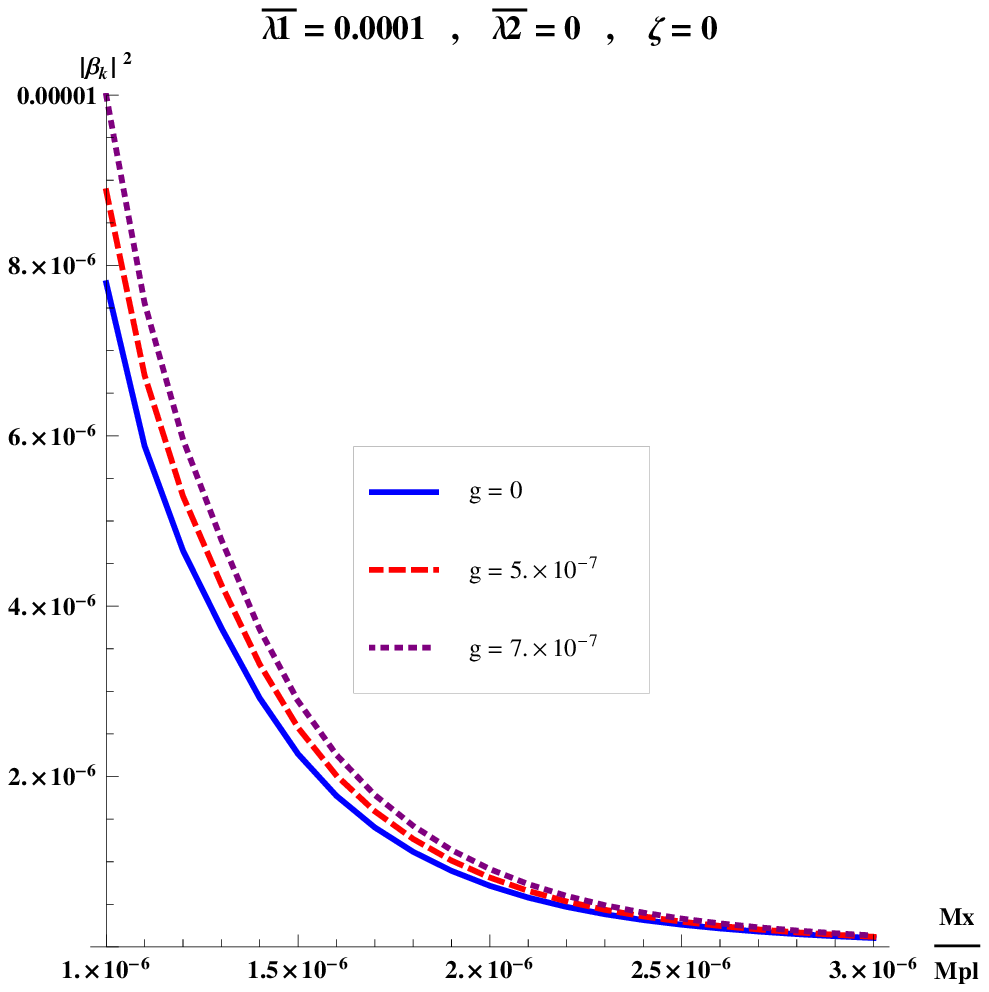}
\caption{Bogolyubov coefficients of the X-field with no couplings
to Einstein tensor, for various values of its coupling to the
inflaton ($\overline{\lambda_1} = 0.0001$).} \label{l0001.eps}
\end{figure}

\begin{figure}[H]
\centering \includegraphics[scale=0.7]{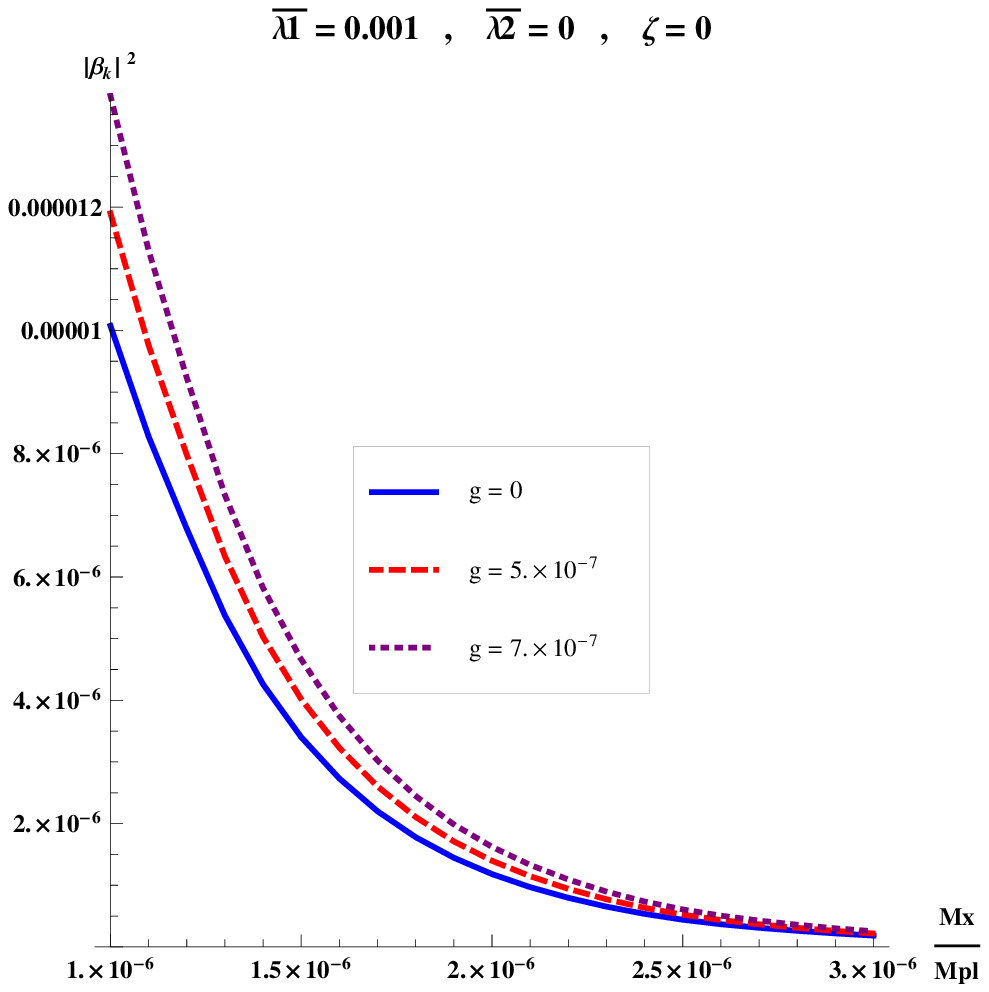}
\caption{Bogolyubov coefficients of the X-field with no couplings
to Einstein tensor, for various values of its coupling to the
inflaton ($\overline{\lambda_1} = 0.001$).} \label{l001.eps}
\end{figure}

\begin{figure}[H]
\centering \includegraphics[scale=0.7]{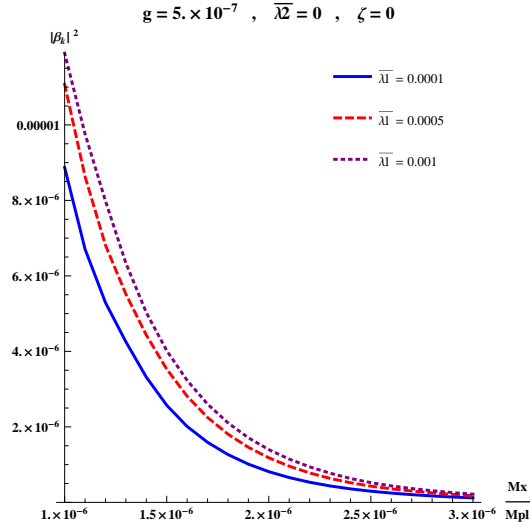}
\caption{Bogolyubov coefficients of the X-field for various couplings  of the Einstein tensor to the
inflaton} \label{ldiaf.eps}
\end{figure}

\begin{figure}[H]
\centering \includegraphics[scale=0.85]{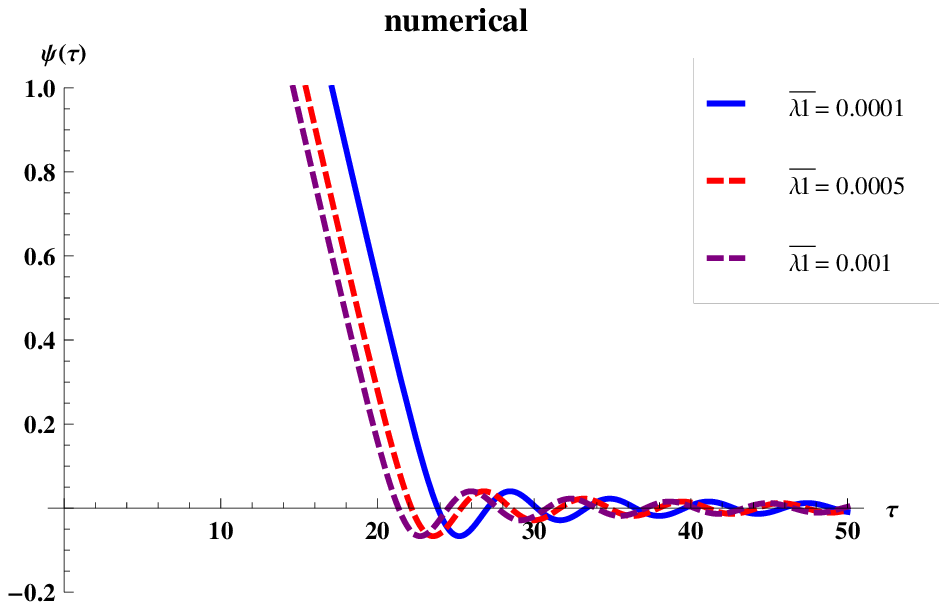}
\caption{Evolution of inflaton $\psi$ as a function of $\tau$ for positive values of the coupling $\overline{\lambda_1}$} \label{tlroll.eps}
\end{figure}

For completeness let us discuss the particle production in the
very small window of positive values for $\overline{\lambda}_1$
given in relation (\ref{rel13}). In Fig. \ref{l0001.eps} and Fig. \ref{l001.eps} we set $\overline{\lambda_2} = 0$ and we show the particle production for the couplings $\overline{\lambda}_1 =0.0001$
and $\overline{\lambda}_1 = 0.001$ respectively. This time we notice an enchancement on the particles produced. We observe the same thing if we set $g=5\cdot 10^{-7}$ and draw the graph for different values of $\overline{\lambda_1}$ (Fig. \ref{ldiaf.eps}) .  However as we already stated the window of positive values for $\overline{\lambda_1}$ is small if one wants to have a proper inflation. This can be checked in Fig. \ref{tlroll.eps} were we show that by increasing the value of a positive $\overline{\lambda_1}$ one ends up with a more steep roll of the inflaton and questions of instability are arising.

In summary, the introduction of the coupling to Einstein tensor of
the inflaton field or of the X-field results in the suppression of
the number of particles produced after the end of  inflation.

\section{Conclusions}

We have studied the particle production due to the expansion of the
Universe of a scalar field coupled to Einstein tensor. After
reviewing the mechanism of gravitational particle production we
applied it to a FRW expanding Universe. Introducing a coupling of a
scalar field to the Einstein tensor we find that the number of
gravitationally produced particles is decreasing as the strength of
the coupling to the Einstein tensor is increasing.

In a more realistic setup, we studied  the particle production due
to the expansion of the Universe after the end of inflation in the
preheating epoch. Introducing a X-field coupled to the inflaton
field we find that as the strength of the coupling is increased the
number of X-particles produced is increased, as expected.

After reviewing the inflationary phase driven by an inflaton field
coupled to the Einstein tensor we  introduced a coupling of the
X-field to Einstein tensor. We carried out a detailed study of
gravitationally produced heavy X-particles in the presence of the
derivative couplings. We found that as the strength of the couplings
of either the inflaton field or the X-particles to the Einstein
tensor is increased, less particles are produced. The dominant
effect comes from the coupling of the inflaton to the Einstein
tensor as the Einstein tensor  after the end of inflation attenuates
due to the expansion of the Universe.

As we  discussed the presence of the coupling of the scalar field
to Einstein tensor acts as a friction term absorbing energy from
the kinetic energy of the scalar field. We can attribute the
suppression of the particle production to the same mechanism. As
the strength of the coupling $\lambda_1$ is increased, less energy
is transferred to the X-particles through the coupling of the
inflaton field to the X-field, so less particles are produced.
This can also be understood as an  effect of the geometry.
Curvature effects are strong during inflation absorbing energy,
while after inflation curvature is small and this is the reason
why the coupling of the X-field to Einstein tensor does not give
sizeable effects.

The main theoretical problem of particle production after the end
of inflation is to control the number of particles produced in
such a way as not to overclose the  Universe. This is achieved
with a fine-tuning of parameters. The gravitational particle
production is a dynamical mechanism between the classical
gravitational field and a quantum field. For this reason has less
fine-tuning of parameters. However, to have the right number of
particles produced, certain assumptions should be fulfilled. It
would be interesting to apply this suppression mechanism we
discussed in this work to produce a realistic cosmological model
of particle production after the end of inflation.

\section*{Acknowledgements}
We would like to thank M. Saridakis and S. Tsujikawa for valuable
discussions and correspondence.

\end{document}